\documentclass[9pt]{article}
\usepackage[utf8]{inputenc}
\usepackage[T1]{fontenc}
\usepackage{mathpazo}
\usepackage{amsmath}
\usepackage{amssymb}
\usepackage[letterpaper, top=1.8cm, bottom=1.8cm, left=1.3cm, right=1.3cm, columnsep=0.6cm]{geometry}
\usepackage{caption}
\usepackage{tikz}
\usepackage{lineno}   
\usepackage[shortcuts]{extdash}  
\usepackage{float}
\usepackage{titlesec}
\usepackage{changepage}
\usepackage{hyperref}

\hypersetup{
    colorlinks=true,
    linkcolor=blue,
    citecolor=blue,
    urlcolor=blue
}

\titleformat*{\section}{\fontsize{11}{12}\selectfont\bfseries}

\titleformat*{\subsection}{\fontsize{9}{12}\selectfont\bfseries}

\begin{document}


\begin{center}
    {\LARGE \textbf{A substrate booster for P-type 2D ferromagnetic semiconductor}}\\[1.3em]
    
    { Hai Wang,$^{1,2*}$ Woye Pei,$^{3,4*}$ Ridong Cong,$^{5*\dagger}$ Xiaoyan Liu,$^{5}$ Yuping Tian,$^{6}$ Kenji Watanabe,$^{7}$ Teng Yang,$^{2,8}$  Takashi Taniguchi,$^{9}$ Xiangru Kong,$^{6}$ Weijiang Gong,$^{6}$ Guowei Zhou,$^{10}$ Xiaoxi Li,$^{3,4}$ Hanwen Wang,$^{2}$ Jixuan Wu,$^{1\dagger}$  Jiezhi Chen,$^{1\dagger}$  Xiaohong Xu,$^{10\dagger}$ Tongyao Zhang$^{3,4,2\dagger}$}
\\[1.5em]
    
    \footnotesize
{
    $^{1}$ School of Information Science and Engineering, Shandong University,  Qingdao, China
\\
    $^{2}$ Liaoning Academy of Materials, Shenyang, China\\
    $^{3}$ State Key Laboratory of Quantum Optics and Quantum Optics Devices, Institute of Optoelectronics, Shanxi University, Taiyuan, China\\
    $^{4}$ Collaborative Innovation Center of Extreme Optics, Shanxi University, Taiyuan, China\\
    $^{5}$ National and Local Joint Engineering Laboratory of New Energy Photoelectric Devices, College of Physics Science \& Technology, Hebei University, Baoding, China\\
$^{6}$ College of Sciences, Northeastern University, Shenyang, China \\
$^{7}$ Research Center for Electronic and Optical Materials, National Institute for Materials Science, Tsukuba, Japan \\
$^{8}$ Shenyang National Laboratory for Materials Science, Institute of Metal Research, Chinese Academy of Sciences, Shenyang, China\\
$^{9}$ Research Center for Materials Nanoarchitectonics, National Institute for Materials Science, Tsukuba, Japan\\
$^{10}$ Key Laboratory of Magnetic Molecules and Magnetic Information Materials of Ministry of Education, School of Materials Science and Engineering, Shanxi Normal University, Taiyuan, China \\[1.2em]
}
    
    $^{\dagger}$ Corresponding author: \href{congrd@hbu.edu.cn},{jixuanwu@sdu.edu.cn},  { chen.jiezhi@sdu.edu.cn }, {xuxh@sxnu.edu.cn}, {tongyao\_zhang@sxu.edu.cn}\\
    $^{*}$ These authors contributed equally to this work.\\[2.5em]
\end{center}

\begin{adjustwidth}{1cm}{1cm}
\noindent \textbf{Abstract}\\[0.5em]
\small
\textbf{
Spin transistors with its both charge and spin properties tuned via electrostatic gating are believed capable for widespread use, which however have proven challenging due to the extreme rareness of their physical base -- magnetic semiconductors. The latter are limited within very few systems including diluted magnetic semiconductors (DMS) and two-dimensional ferromagnetic semiconductors (2D-FMS), and known to suffer from inadequate gate-tunability of their electric and/or magnetic properties. Here, we show a substrate engineering paradigm by interfacing few-layered Cr$_{2}$Ge$_{2}$Te$_{6}$ (FL-CGT) with an antiferromagnetic insulator CrOCl. Owing to the subtle interfacial charge transfer couplings, CGT can be drastically turned from an ambipolar semiconductor into a high performance P-type semiconductor. When cooled below the Curie temperature, the ON-OFF ratio in such substrate-boosted FMS field-effect transistor (FET) reaches 10$^{5}$ with its coercive field $H_{c}$ of magnetic hysteresis loop tunable by a factor of more than 200$\%$, enabling  {gate-assisted magnetic switching in the prototype semiconducting spin transistor architecture}. A crossover from critical power-law scaling to a dual power-law behaviour under heavy hole doping was further observed. Our findings  {signify} an efficient interfacial charge transfer and electrically modulated magnetic anisotropy energy supported by calculations. This high performance P-type FMS-FET system suggests that active substrate-boosting paradigm might be a powerful path for the investigation of future gate-tunable spintronic devices.
}
\end{adjustwidth}

\clearpage

\twocolumn

\section*{Introduction}

Magnetic semiconductors that function akin to the well\-/developed semiconducting materials, with both spin- and charge-properties tunable by electrical fields, have been a dream system since years. Spin transistors, in which both charge and spin can be modulated electrically, are widely regarded as a central building block for future low-power, non-volatile and multifunctional devices\cite{datta1990electronic,vzutic2004spintronics}. However, their practical realization critically relies on magnetic semiconductors that simultaneously support long-range magnetic order and gate-tunable carrier transport. Such materials are exceedingly rare in nature, as magnetism and semiconducting behaviour are often mutually exclusive\cite{dietl2002ferromagnetic,li2016first}. To date, experimentally accessible systems remain largely confined to diluted magnetic semiconductors\cite{ohno2000electric,matsumoto2001room,chiba2003electrical,wang2020high} and a limited family of two\-/dimensional ferromagnetic semiconductors\cite{lin2016ultrathin,Wang2018a,jiang2018controlling,zhao2025doping}, constraining both device performance and scalability.

Despite intensive efforts, existing ferromagnetic semiconductors suffer from fundamental challenges that hinder their integration into functional spin transistors. In particular, weak electrostatic control over carrier density, poor on-state conductivity, and limited tunability of magnetic properties have persistently restricted device operation. These limitations stem from the intrinsic electronic structure of magnetic semiconductors, in which magnetic exchange, band alignment, and carrier concentration are strongly intertwined. Overcoming this bottleneck requires a strategy that enhances charge transport and magnetic response without destroying ferromagnetic order, ideally through a non-invasive and reversible route compatible with field-effect architectures\cite{paul2023giant}. Interfacial engineering, especially via van der Waals (vdW)  heterostructures,  {offers such a pathway by enabling subtle charge redistribution, orbital hybridization, and topological structures while preserving the host lattice and magnetic ground state\cite{zhong2017van,zhong2020layer,choi2022emergent,zhang2022electrically,zhang2024spin}}.

Here we demonstrate that surface contact with chromium Oxychloride (CrOCl) can dramatically reshape the electronic and magnetic properties of FL-CGT. Interfacial charge-transfer coupling converts Cr$_{2}$Ge$_{2}$Te$_{6}$  from an ambipolar semiconductor to a high-performance unipolar P-type ferromagnetic semiconductor, yielding field-effect transistors with on\-/state currents up to 10 $ \mu$A per square and ON-OFF ratios exceeding 10$^{5}$ at cryogenic temperatures. Below the Curie temperature, the coercive field of the magnetic hysteresis loop read by magneto-optic Kerr effect (MOKE) becomes strongly gate-tunable, varying by more than 200\% across the electrostatically accessible regimes, and enabling deterministic switching of magnetization  {assisted by electric\-/field}. Further thermodynamic analysis reveals a crossover from a singular power\-/law critical scaling to a dual power law under extreme hole-doping condition. Although the operation temperature remains below room temperature, these results establish active interfacial substrate boosting as a powerful and general strategy for realizing gate-tunable spintronic devices based on ferromagnetic semiconductors.

\section*{Results and Discussion}

\subsection*{Fabrications and characterizations of Cr$_{2}$Ge$_{2}$Te$_{6}$ FETs.} 

CrOCl is a prominent stripy antiferromagnet characterized by an orthogonal lattice and a N\'eel temperature of $T_N\sim 13$ K\cite{angelkort2009observation}. Beyond its application in spintronic prototypes leveraging its magnetic order\cite{zhang2022tuning,gu2023multi,cao2025magnetic}, CrOCl has recently emerged as a powerful constituent in "charge-transferonic" heterostructures. When interfaced with materials such as graphene\cite{wang2022quantum,yang2023unconventional,lu2023synergistic} or MoS$_2$\cite{guo2024van}, CrOCl acts as a highly efficient hole reservoir, profoundly modulating carrier dynamics by depleting electrons from the adjacent conducting layer. Recent theoretical models suggest that such charge transfer may induce long-wavelength charge order under specific electrostatic gating. Though direct spectroscopic verification remains elusive, the ability of CrOCl to facilitate complementary metal-oxide-semiconductor (CMOS) inverters in Bernal bilayer graphene\cite{yang2023unconventional} and MoS$_2$\cite{guo2024van} underscores its versatility as a functional substrate. Despite these advances, the potential for CrOCl to modulate the electronic, and/or even magnetic properties of 2D-FMS, such as CGT, remains an open and critical question.

\begin{figure*}[ht!]
 	\centering
 	\includegraphics[width=0.9\linewidth]{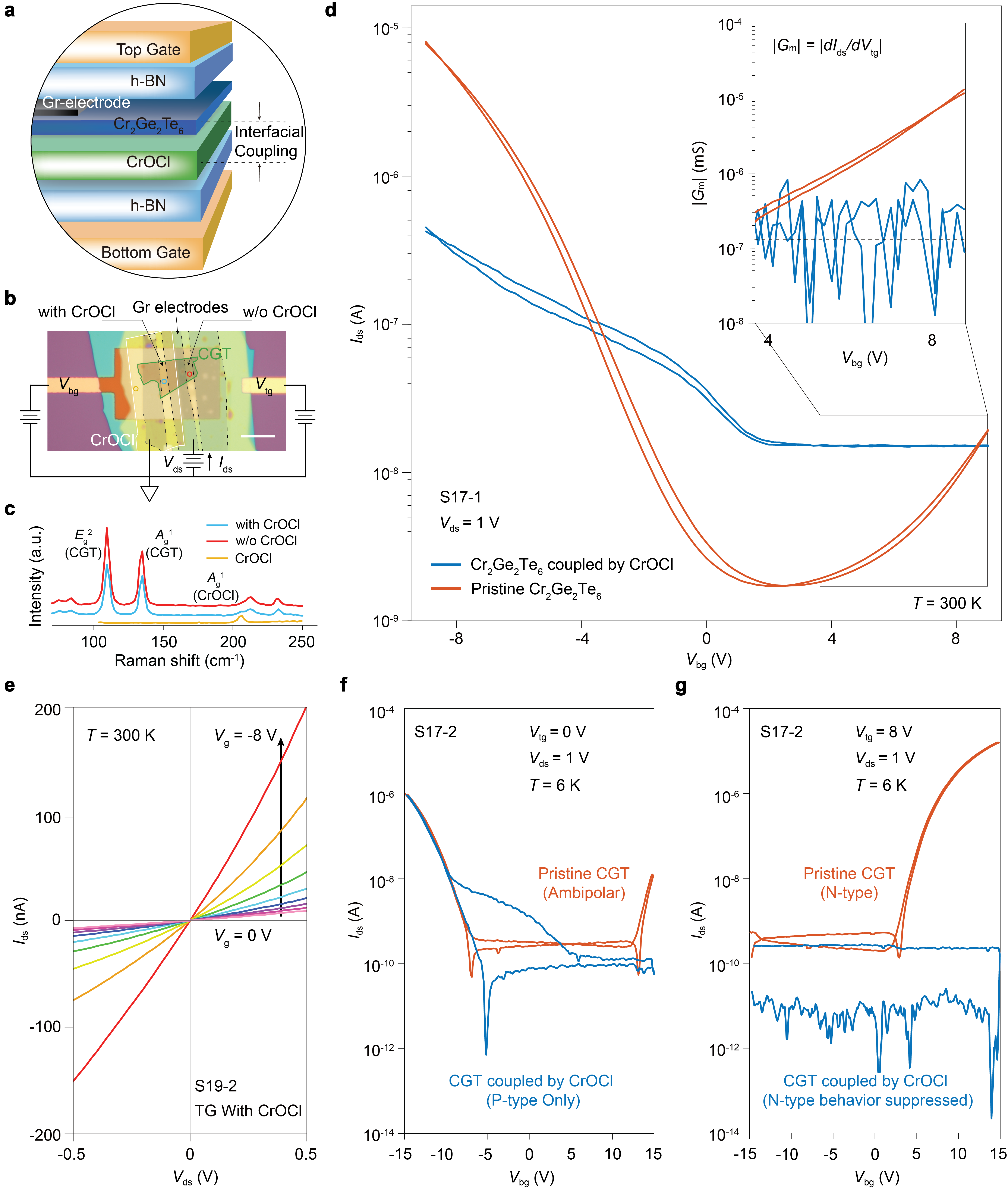}
 	\caption{
    \textbf{Characterization of CrOCl-coupled FL-CGT FETs.} 
    \textbf{a}, Schematic illustration of the dual-gated FET structure consisting of FL-CGT interfaced with a CrOCl underlayer. 
    \textbf{b},  {Optical image of a typical device and the configuration of measurement circuits.} Solid outlines indicate the FL-CGT (green) and CrOCl (white) flakes, dashed black lines mean graphite electrodes. Scale bar, 10 $ \mu$m. 
    \textbf{c}, Raman spectra collected from the distinct regions marked by colored circles in (\textbf{b}). 
    \textbf{d}, Room-temperature transfer characteristics $I_\text{ds}$-$V_\text{bg}$ of device S17-1, comparing the bare FL-CGT channel (red) with the CrOCl-coupled region (blue) at $V_\text{ds}=$ 1 V. The inset displays the magnitude of the transconductance $(G_m)$ in the electron doping regime, highlighting the suppression of N-type transport by the CrOCl layer, which drives the FET into a purely p-type.
    \textbf{e}, Output characteristics $I_\text{ds}$-$V_\text{ds}$ of the CrOCl-coupled channel in a gate range from -8 V to 0 V, exhibiting  {quasi-ohmic at low bias and nonlinear behavior at high bias}.
    \textbf{f}, \textbf{g}, Transfer curves at $T=$ 6 K with the top gate set to $V_{\text{tg}}=0 $ V (\textbf{f}) and $V_{\text{tg}}=8 $ V (\textbf{g}), respectively. While the uncoupled FL-CGT region shows ambipolar (\textbf{f}) or N-type behavior (\textbf{g}), the CrOCl-coupled channel remains p-type (\textbf{f}) or being fully depleted (\textbf{g}) under positive gating.
    }
	\label{fig:fig1}
 \end{figure*}

To address this, we propose and fabricate dual-gated FETs based on CrOCl-coupled FL-CGT using a hybridized dry-transfer technique (Fig.~\ref{fig:fig1}a; see Methods and \textcolor{black}{Supplementary Figure 1} for details). To prevent ambient degradation and provide atomically flat interface, FL-CGT flakes ($<5$ nm) were encapsulated between h-BN dielectrics with thickness $\sim 20$--$40$ nm (\textcolor{black}{Supplementary Figure 2}). For most of the devices in the main text, the intercalated CrOCl layer is positioned beneath the FL-CGT channel; we confirmed that a mirror-image configuration (top-layer CrOCl, \textcolor{black}{Supplementary Figure 3}) yields consistent results (\textcolor{black}{Supplementary Figure 4}). To {improve} ohmic contacts, exfoliated graphite flakes were employed as electrodes. Fig.~\ref{fig:fig1}b displays an optical micrograph of a representative device, where the stacking of FL-CGT/CrOCl is strategically designed to create two distinct regions: one interfaced with CrOCl and one pristine. Micro-Raman spectroscopy (Fig.~\ref{fig:fig1}c) confirms the structural integrity of the heterostructure (see Methods). {The characteristic $E_{g}^{2}$ and $A_{g}^{1}$ modes persist in the coupled region (blue spectrum), alongside the CrOCl $A_{g}^{1}$ mode at $\sim 206 $ cm$^{-1}$, demonstrating clean vdW stacking without lattice degradation}.

The impact of this coupling on electrical transport is evaluated via direct-current measurements. Fig.~\ref{fig:fig1}d compares the transfer characteristics of pristine FL-CGT and CrOCl-coupled CGT at $T = 300$ K ($V_{\text{ds}} = 1$ V). While the pristine FET exhibits the expected ambipolar behavior, the CrOCl-interfaced device shows a suppressed hole-side $I_{\text{ds}}$, partially attributed to the reduced effective capacitance of the bottom gate, and a complete suppression of electron transport in the positive $V_{\text{bg}}$ regime. This nearly constant off-state conductance is further evidenced by the vanishing transconductance ($|G_{\text{m}}| = |dI_{\text{ds}}/dV_{\text{bg}}|$) in the electron-doping range (inset of Fig.~\ref{fig:fig1}d). Although the output characteristics remain linear {under low bias} (Fig.~\ref{fig:fig1}e), the incomplete depletion by CrOCl {and finite Schottky barrier under high bias} at room temperature limits the ON-OFF ratio of the device. However, upon cooling to $T = 6$ K within the CGT ferromagnetic phase, the CrOCl-coupled FET achieves an ON-OFF ratio exceeding $10^{5}$, a full order of magnitude higher than the $10^{4}$ observed in pristine CGT (Fig.~\ref{fig:fig1}f). Due to the Schottky $I_{\text{ds}}-V_{\text{ds}}$ response at low temperature, the on-state $I_{\text{ds}}$ could reach 10 $\mu$A/$\square$ at $V_{\text{ds}} = 2$ V (\textcolor{black}{Supplementary Figure 5}). Despite a slight hysteresis indicating interfacial disorder-induced trapping, the transition from ambipolar to strictly P-type behavior demonstrates that CrOCl acts as an exceptionally effective electrostatic depletion for electrons. { This behavior is governed by the interfacial band alignment that drives spontaneous electron transferring from CGT into CrOCl, and is further corroborated by Kelvin probe force microscopy (KPFM) measurements (Supplementary Figure 6) confirming the up-shift electrostatic potential of FL-CGT.} Even under substantial electron injection from the top gate ($V_{\text{tg}} = 8$ V), the powerful hole-reservoir effect of the CrOCl interface maintains the device in the off state (Fig.~\ref{fig:fig1}g).

\subsection*{Boosted performance of gate tunable coercivity fields in Cr$_{2}$Ge$_{2}$Te$_{6}$ FETs.} 

\begin{figure*}[ht!]
\centering
\includegraphics[width=0.9\linewidth]{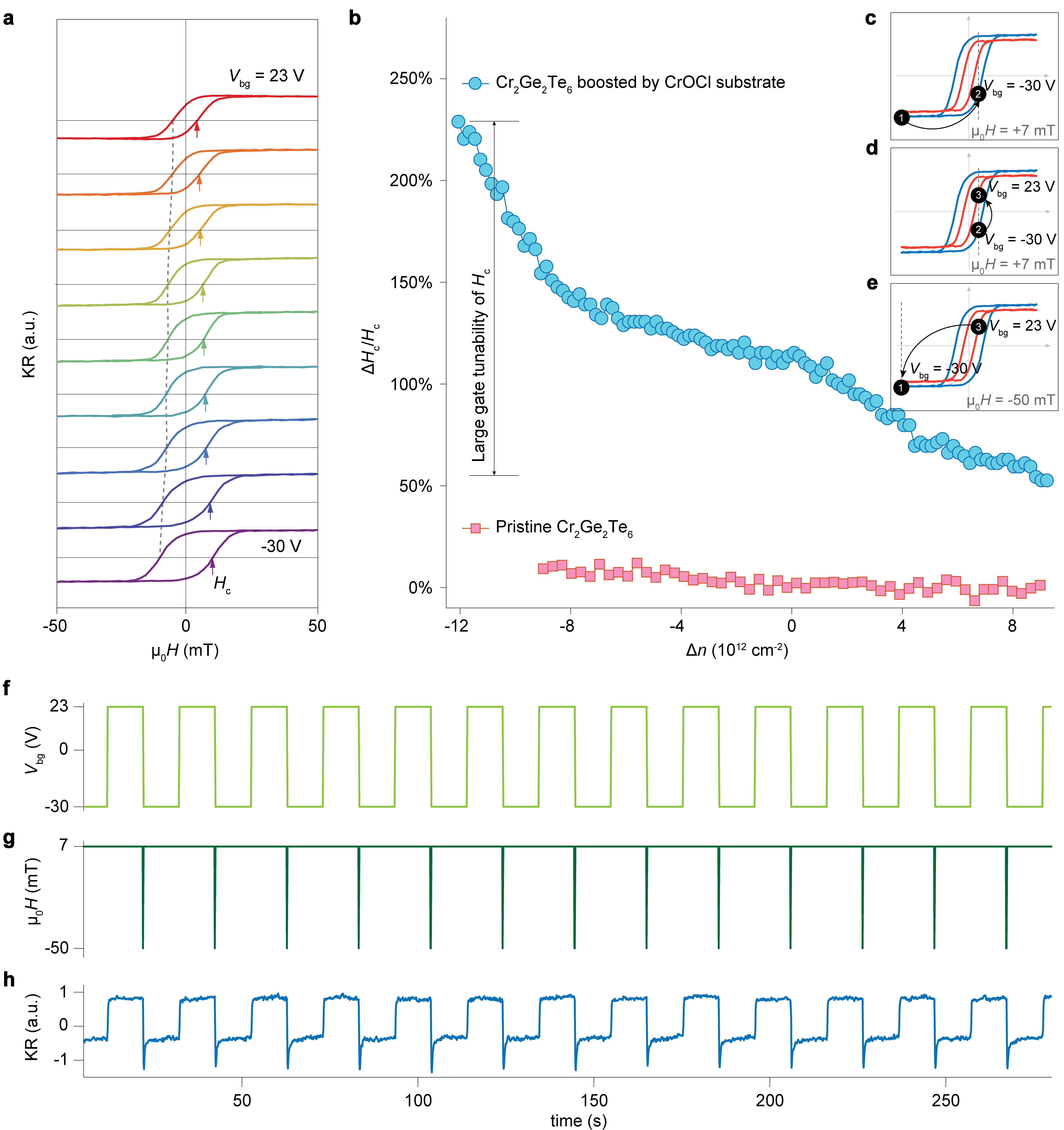}
\caption{
\textbf{Giant gate-tunability of coercivity fields in CrOCl-coupled FL-CGT devices.} 
\textbf{a}, Vertical stacking of magnetization hysteresis loops measured by MOKE from $V_{\text{bg}}=-30 $ V to $V_{\text{bg}}=23 $ V at 6 K. The set gate $V_{\text{tg}}=-14 $ V. A strong dependence of the coercive field ($H_{\text{c}}$) on the electric field is observed.
\textbf{b}, Relative change in coercivity $\Delta H_\text{c}/H_\text{c}$ versus electrostatic doping $\Delta n$. The CrOCl-coupled device (blue circles) exhibits a giant tunability of over 200\%, significantly exceeding the negligible response of the pristine CGT device (pink squares). Data are normalized to the minimum $H_\text{c}$ observed in the dual-gate map (\textcolor{black}{Supplementary Figures 8}). 
\textbf{c} - \textbf{e}, Schematic illustration of the electrically {assisted} magnetic switching protocol. 
\textbf{c}, Applying a constant bias field $\mu_0H=7$ mT and set $V_\text{bg}=-30$ V to an initial state 2 ("spin-down"). 
\textbf{d}, Switching the gate from state 2 to $V_\text{bg}=23$ V (state 3) reverses the magnetization to "spin-up". 
\textbf{e}, The magnetization is reset to state 1 by a $-50$ mT magnetic pulse to prepare for next circle.
\textbf{f}, Demonstration of deterministic electrical writing and optical reading. The KR (bottom) tracks the alternation of $V_\text{bg}$ between -30 V and 23 V (top), confirming reversible magnetization switching assisted by a constant bias field (7 mT) and pulsed reset fields (middle). 
}
\label{fig:fig2}
\end{figure*}

To elucidate the influence of the CrOCl substrate on the magnetic properties of FL-CGT, we performed polar MOKE microscopy to monitor the magnetization hysteresis loops under dual-gate control at $T = 6$ K with zero field cooling (see Methods). The magnetic uniformity of our as-fabricated devices is confirmed by spatial mapping of Kerr rotation (KR) hysteresis (\textcolor{black}{Supplementary Figure 7}). By systematically mapping the dual gate dependence, we find that the coercive field $H_{\text{c}}$ and residue magnetization over saturation magnetization $M_r/M_s$ extracted from the hysteresis loops monotonically increase with doping holes (\textcolor{black}{Supplementary Figure 8}), growing to be harder ferromagnet. Furthermore, we select a regime (fixed $V_{\text{tg}} = -14$ V) where the KR loops exhibit a dramatic response to the bottom gate voltage ($V_{\text{bg}}$ ranging from $-30$\,V to $23$\,V). As shown in Fig.~\ref{fig:fig2}a, $H_{\text{c}}$ is profoundly modulated by the applied electric field. The saturation magnetization remains largely invariant (\textcolor{black}{Supplementary Figure 8}), suggesting that the electrostatic modulation primarily tunes the magnetic anisotropy energy, rather than the total magnetic moment. The relative enhancement of the coercive field, defined as $\Delta H_{\text{c}}/H_{\text{c}} = (H_{\text{c}} - \min(H_{\text{c}}))/\min(H_{\text{c}})\times100\%$ where $\min(H_{\text{c}})$ is the minimum of the dual-gate mapping, is plotted against the estimated electrostatic doping concentration $\Delta n$ in Fig.~\ref{fig:fig2}b. {$\Delta n$ is estimated to be $\Delta n = C_{\text{bg}}V_{\text{bg}}/e$, where $C_{\text{bg}}=\varepsilon_0 \varepsilon_{\text{hBN}}\varepsilon_{\text{CrOCl}}/(\varepsilon_{\text{CrOCl}}d_{\text{hBN}} + \varepsilon_{\text{hBN}}d_{\text{CrOCl}})$, $e$ is the element charge, $\varepsilon_0$ is the vacuum permittivity, $\varepsilon_{\text{hBN}}, \varepsilon_{\text{CrOCl}}, d_{\text{hBN}}, d_{\text{CrOCl}}$ are relative permittivity and thickness of bottom h-BN and CrOCl layer respectively } In stark contrast to a controlled pristine FL-CGT sample (magenta squares, S14-2), which shows a modest $\sim 10\%$ increase in $H_{\text{c}}$ {from $\sim$ 6.0 mT to $\sim$ 6.6 mT} under electron doping, the CrOCl-coupled device ( {cyan} circles, S12-5) exhibits a giant enhancement exceeding $230\%$ {from $\sim$ 3.1 mT to over 10.0 mT} in the hole-doped regime. Such giant tunability of $H_\text{c}$ is observed in an individual sample S3-1 (\textcolor{black}{Supplementary Figure 8}). Furthermore, the magnetic energy product increase under hole doping, which peaks at the negative limits of our dual-gate range (\textcolor{black}{Supplementary Figure 9}). The magnetic properties tunability versus dual gate of an individual pristine FL-CGT sample is shown in (\textcolor{black}{Supplementary Figure 10}) for comparison. This unprecedented tunability in a 2D-FMS implies that the CrOCl interface significantly bolsters the magnetocrystalline anisotropy energy (MAE)). Notably, our platform implies the electrical reversal ability of the KR polarity at a finite external magnetic field, a capability previously elusive in FL-CGT systems.

Leveraging this gate-dependent magnetic polarity, we propose and demonstrate a {gate-assisted} magnetic switching protocol characterized by electrical writing and optical reading (Fig.~\ref{fig:fig2}c--e). The device is initialized into a "spin-down" state (State 2) by sweeping the magnetic field from $-50$\,mT (state 1) to $7$\,mT at $V_{\text{bg}} = -30$ V (Fig.~\ref{fig:fig2}c). Maintaining a constant bias field of $\mu_0 H = 7$\,mT, the KR is smoothly transitioned to a positive "spin-up" state (State 3) by tuning the gate to $V_{\text{bg}} = 23$ V (Fig.~\ref{fig:fig2}d). Finally, the magnetization is fully reversed to State 1 via a magnetic field pulse (Fig.~\ref{fig:fig2}e). Throughout this cycle, the top gate remains constant at $V_{\text{tg}} = -14$ V, with the switching driven primarily by the bottom gate in the presence of a moderate bias field. A typical real-time demonstration of this deterministic magnetization switching is presented in Fig.~\ref{fig:fig2}f. By cyclically alternating $V_{\text{bg}}$ between $-30$\,V and $23$\,V, we observe reversible and reproducible switching of the magnetization states. Crucially, this manipulation is driven by electrostatic field effects rather than dissipative charge currents, highlighting the potential of CrOCl-coupled FL-CGT for ultra-low-power spintronic applications.

\subsection*{Temperature-dependent gate modulation of magnetic properties in CrOCl-coupled Cr$_{2}$Ge$_{2}$Te$_{6}$.} 

\begin{figure*}[ht!]
\centering
\includegraphics[width=0.85\linewidth]{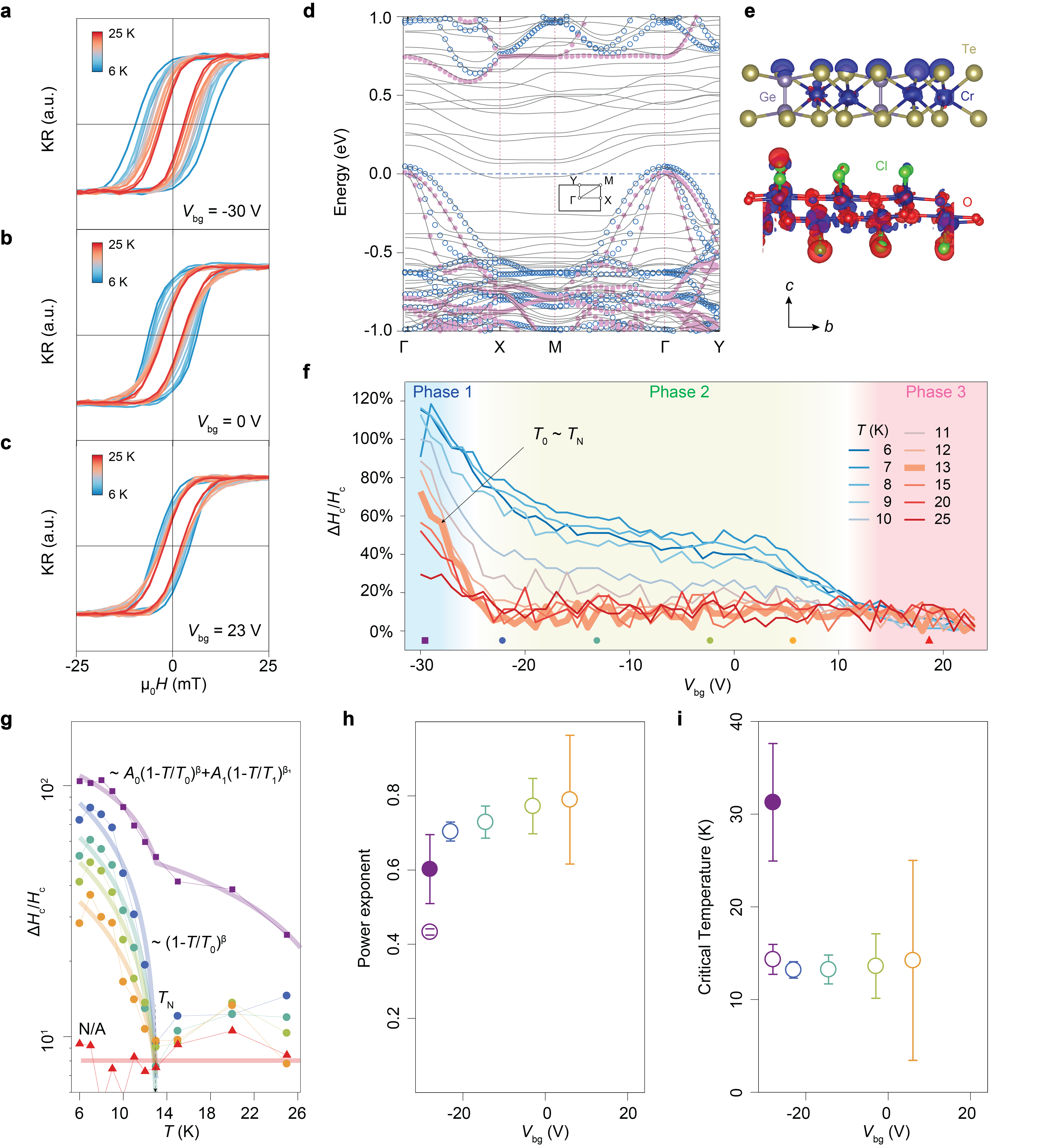}
\caption{
\textbf{Temperature-dependent gate modulation of magnetic coercivity in CGT/CrOCl heterostructures.} 
\textbf{a}--\textbf{c}, Temperature dependence of normalized KR hysteresis loops from 6 K to 25 K at $V_{\text{bg}} = -30$ V (\textbf{a}), 0 V (\textbf{b}), and 23 V (\textbf{c}), respectively. The top-gate voltage is fixed at $V_{\text{tg}} = -14$ V. 
{\textbf{a-c} share the temperature color scale defined in \textbf{f}.} 
\textbf{d}, Electronic band structure of CGT/CrOCl hybrid (black lines). Solid purple and empty blue circles represent the spin majority and minority electronic bands of CGT monolayer, respectively. 
\textbf{e}, Differential charge density $\Delta \rho = \rho($CGT/CrOCl$) - \rho($CGT$) - \rho($CrOCl$)$. Blue and red colors represent charge depletion and accumulation, respectively. CGT (top layer) is positively charged while CrOCl (bottom layer) is negatively charged. 
\textbf{f}, Relative coercivity fields ($\Delta H_{\text{c}}/H_{\text{c}}$) as a function of $V_{\text{bg}}$ across the 6-25 K range. Values are normalized to the coercivity minimum of each respective curve. The response is partitioned into three distinct phases: Phase 1 hole-doped regime (blue), characterized by significantly enhanced coercivity tunability, Phase 2 moderate doping regime (light yellow), and Phase 3 an electron-depletion regime (red), where the proximity-induced enhancement vanishes. Above the Néel temperature of CrOCl, $\Delta H_{\text{c}}/H_{\text{c}}$ converges toward the experimental noise floor.
\textbf{g}, Temperature dependence of $\Delta H_{\text{c}}/H_{\text{c}}$ at selected $V_{\text{bg}}$ (indicated by symbols in (\textbf{f})). In the moderate doping regime, the data follow a power-law scaling, $\Delta H_{\text{c}}/H_{\text{c}} \propto (1 - T/T_0)^{\beta}$ (thick lines). In the heavily hole-doped regime (purple), the decay follows a dual power law $A_0(1-T/T_0)^\beta+A_1(1-T/T_1)^{\beta_1}$ (purple thick line), suggesting a second {characteristic} temperature of $T_1\approx 31$ K. Negligible signal is observed in the depletion regime (red triangles). 
\textbf{h}, \textbf{i}, Extracted critical exponent $\beta$ (open circles) (\textbf{h}), critical temperature $T_0$ (open circles) (\textbf{i}) from the power-law fits in (\textbf{e}). The parameters $\beta_1$ and $T_1$ revealed from the second component of the dual power law are plotted in solid circles. Error bars represent the fitting errors. 
} 
\label{fig:fig3}
\end{figure*}

To further study the physical origin of the CrOCl-enhanced magnetization modulation, we investigate the thermal stability of the CrOCl-coupled system. Figs.~\ref{fig:fig3}a--c depict the temperature-dependent KR hysteresis loops for device S12-5 under three typical gate conditions. As expected, the hysteresis loops undergo a monotonic contraction with increasing temperature across all $V_{\text{bg}}$ values, consistent with the escalation of thermal fluctuations. At any given temperature, the loops narrow as $V_{\text{bg}}$ is swept from the hole-doping to the electron-doping regime, maintaining the trend observed at base temperature (Fig.~\ref{fig:fig2}a). 

Following the procedures for base-temperature analysis to eliminate the background of pristine CGT magnetic properties (\textcolor{black}{Supplementary Figure 11}), we calculated the relative change in coercivity, $\Delta H_{\text{c}}/H_{\text{c}}$, normalized to the minimum $H_{\text{c}}$ at each respective temperature (Fig.~\ref{fig:fig3}f). Note that for these analysis, $V_{\text{tg}}$ is fixed at $-14$ V , resulting {in} a baseline downward shift compared with the data in Fig.~\ref{fig:fig2}b. Similar temperature-dependent behavior in an additional device is provided in \textcolor{black}{Supplementary Figure 12}. As expected, $\Delta H_{\text{c}}/H_{\text{c}}$ increases with hole doping across all studied temperatures. However, the gate tunability of $H_{\text{c}}$ weakens and the threshold $V_{\text{bg}}$ shifts toward higher values upon warming. This suggests that the interfacial coupling between CrOCl and FL-CGT is highly sensitive to the magnetic order of the substrate. Notably, $\Delta H_{\text{c}}/H_{\text{c}}$ drops toward the noise floor above $T \approx 13$ K, a value nearly coincident with the N\'eel temperature of CrOCl. 

To further quantitatively characterize the gate-tunable magnetic coupling, we analyze the temperature dependence of $\Delta H_{\text{c}}/H_{\text{c}}$ at several $V_{\text{bg}}$ (markers at the bottom of Fig.~\ref{fig:fig3}f). As shown in Fig.~\ref{fig:fig3}g, critical scaling analysis reveal three distinct regimes. In the moderate doping regime, (Phase 2, -23 V < $V_{\text{bg}}$ < 8 V ), $\Delta H_{\text{c}}/H_{\text{c}}$ could be well interpreted by a power-law scaling, $\Delta H_{\text{c}}/H_{\text{c}}\propto (1-T/T_0)^\beta$, extracting critical component $\beta$ (Fig.~\ref{fig:fig3}h) and characteristic temperature $T_0$ (Fig.~\ref{fig:fig3}i), respectively. The $T_0$ in Fig.~\ref{fig:fig3}i is consistent to $T_N$ of CrOCl within the error range, {indicating an interfacial coupling linked to the antiferromagnetic order of substrate}. The scaling exponent $\beta\approx0.7$ deviates from the mean-field value 0.5, likely reflecting the 2D nature of the interfacial correlations. Such scaling behaviour breaks down in the electron-depletion regime (Phase 3, $V_{\text{bg}}$ > 8 V), where $\Delta H_{\text{c}}/H_{\text{c}}$ collapses into the noise floor (red triangles in Fig.~\ref{fig:fig3}g), indicating the close of substrate boosting effect. 

Intriguingly, in the heavily hole-doped regime ( {Phase 1}, $V_{\text{bg}} < -25~\text{V}$), a divergence of the scaling exponent is observed, {indicating a breakdown of the single-component critical scaling}. Instead, the temperature dependence of $\Delta H_{\text{c}}/H_{\text{c}}$ in this regime is better described by a {phenomenological} dual power law $A_0(1-T/T_0)^\beta+A_1(1-T/T_1)^{\beta_1}$ (purple line in Fig.~\ref{fig:fig3}g). The extracted {characteristic} temperatures clearly reveal two distinct interfacial coupling components: a low-temperature component with $T_0\approx14 $ K, matching the $T_N$ of CrOCl, and an emergent component with $T_1\approx31$ K. 

Crucially, the invariable nature of both $\beta$ and $T_0 \approx T_N$ indicates that the interfacial exchange coupling strength remains largely unperturbed by either doping or the applied electric field. To explain the emergence of $T_1$, we propose that the extreme interfacial electric field and massive hole accumulation {potentially drive the system toward} a localized structural or charge-ordered electrostatic transition at the interface under high negative gating. {If such an interfacial reconstruction occurs, it would modify the} local electrostatic environment, allowing the pinning effect to persist above $T_{\text{N}}$ even slightly above the spin density wave state temperature $T_{\text{SDW}} \sim 27$ K of CrOCl. The robust thermal stability suggests that electrostatic gating not only tunes the magnitude of coercivity enhancement but also {demonstrates the potential to modulate the thermal scaling behavior and stability of the interfacial magnetic pinning}. While these results provide a solid phenomenological framework, a comprehensive microscopic description of this thermodynamic stabilization requires further theoretical and spectroscopic investigation {to definitively resolve these hypothesized interfacial states}.

Notably, the observed substrate boosting effect is fundamentally distinct from conventional exchange bias phenomena common in many ferromagnetic/antiferromagnetic heterostructures. In traditional exchange bias systems, the antiferromagnetic layer typically acts as a passive pinning barrier, providing a static unidirectional anisotropy that merely shifts and broadens the hysteresis loop without fundamentally altering the intrinsic magnetic ordering of the ferromagnet, which is hard to tune via electrostatic gating. {In contrast, the CrOCl substrate in our system serves predominantly as an active, electric-field-controlled modulator rather than a purely passive pinning layer.} This is evidenced by the massive gate-tuned $\Delta H_{\text{c}}/H_{\text{c}}$, and is further corroborated by the complete quenching of the coercivity enhancement in the electron-depletion regime Phase 3. {Upon the depletion of accumulated holes, the heterostructure transitions to a soft-ferromagnetic state. However, interfacial coupling with CrOCl substrate provides static exchange pinning, leading to a baseline increase of $H_{\text{c}}$ compared with pristine FL-CGT (Supplementary Figure 7). Conversely, heavy hole injection effectively transforms FL-CGT/CrOCl hybrid into an platform enabling gate-tuned magnetic hardening and high-anisotropy with thermally robust magnetization.} The active interfacial nature of CrOCl also manifests in the thickness dependence of the electrically modulated magnetic properties (\textcolor{black}{Supplementary Figure 13}). The $\Delta H_{\text{c}}/H_{\text{c}}$ and $\Delta(M_r/M_s)/(M_r/M_s)$ drop sharply as the thickness of FL-CGT increases, denoting a short-range physical mechanism originating directly from interfacial charge transfer.

However, we note that in our exemplary FL-CGT/CrOCl system, the {crossover} temperature $T_1$ remains smaller than the intrinsic Curie temperature of CGT, prohibiting an observable enhancement of the macroscopic $T_{\text{C}}$ in Phase 1. As the temperature exceeds $T_1$, the CrOCl transitions into a paramagnetic state, causing the thermal melting of the magnetic order and significantly attenuating the interfacial pinning effect. Due to these mechanisms, the global $T_{\text{c}}$ of the FL-CGT channel remains gate-independent in these CrOCl-coupled devices (\textcolor{black}{Supplementary Figure 14}).

\subsection*{DFT calculations and micromagnetic simulation.} 

To provide a comprehensive microscopic description of the interfacial charge transfer and establish the bridge toward the enhanced magnetic coupling, we performed DFT calculations (see Methods). The calculated electronic band structure of the hybrid (Fig.~\ref{fig:fig3}d) shows pronounced hole pockets on top of valence bands of CGT around the $\Gamma$ point, together with a rather flat electron pocket from X and M points at the bottom of conduction band of CrOCl. This indicates a net electron transfer from CGT to CrOCl, verifying spontaneous hole doping in CGT. It agrees well with the experimental observation of the unipolar P-type transport transition shown in such as Fig.~\ref{fig:fig1}f--g. The calculated differential charge density (Fig.~\ref{fig:fig3}e), defined as $\Delta \rho = \rho($CGT/CrOCl$) - \rho($CGT$) - \rho($CrOCl$)$, reveals electron charge depleted (blue) from the Cr and Te atoms of CGT and accumulated (red) in CrOCl. The hole density in CGT estimated from both band structure and integrated differential charge density is in the order of $10^{13}$ cm$^{-2}$ (\textcolor{black}{Supplementary Figure 15}). In the mean time, The conduction bottom minimum of CrOCl is flat enough, in agreement with previous studies \cite{wang2022quantum,yang2023unconventional,lu2023synergistic}. It may provide possibilities for tuning Coulomb interactions of electrons that are transferred from CGT. 

To have a microscopic view on the CrOCl-boosted giant tunability of the coercivity field in CGT (i.e., over 200\%), we calculated electric-field dependence of MAE. The magnitude of the built-in electric field due to spontaneous charge transfer in the hybrid is approximately $-0.02$ V/\text{\AA}. {Because an exact mapping from macroscopic gate voltages across a multi-layer dielectric heterostructure to the local microscopic electric field involves complex dielectric screening, band offsets, and interfacial potentials, we adopt a highly simplified electrostatic boundary condition for a qualitative analysis.} Considering $V_{\text{bg}}$ is applied uniformly across the whole device, an additional electric field superimposes on the built-in field (for example, $-0.03$ V/\text{\AA} if $-30$ V is applied across a device thickness of 90 nm). Total simulated electric fields $0$, $-0.02$, $-0.05$ V/\text{\AA}, which approximately correspond to $V_{\text{bg}}$ = 23, 0 and $-30$ V in experiment, respectively (Fig.~\ref{fig:fig3}a-\ref{fig:fig3}c), were used in our calculations to compute and compare MAE. As demonstrated in \textcolor{black}{Supplementary Figure 16}, the calculated MAE increases consistently with the increasing magnitude of the negative electric field (from 0 to $-0.05$ V/\text{\AA}), which may arise from the enhanced spin-orbit coupling due to electric field. 

{Extended DFT calculations confirm that other factors, such as pure electrostatic doping, interfacial mechanical strain are physically insufficient to drive the highly reversible, dynamic magnetic tunability. Quantitative analyses isolating these mechanisms are detailed in Supplementary Figures 17 through 18. Consistent with this calculation, control experiments using non-magnetic 1T-TaS$_2$ layer reveal limited gate modulation of coercivity despite heavy hole transfer (Supplementary Figure 19), confirming that electrostatic hole accumulation alone is insufficient to drive the giant $H_{\text{c}}$ tunability.} To substantiate the direct relevance of electric-field dependent MAE to the enhancement of coercivity field, micromagnetic simulations using the OOMMF code~\cite{oommf} (\textcolor{black}{Supplementary Figure {20}}) was performed and show that coercivity field indeed increases with increasing MAE in the hysteresis loops.

These calculations, combined with the experimental results, highlight that the CrOCl substrate facilitates an active substrate boosting effect to enhance the gate tunability of the coercive fields in FL-CGT FETs. First, the CrOCl layer acts as an efficient electron reservoir, effectively shifting the Fermi level of the FL-CGT channel. As the system cools, the improved band alignment between CrOCl and FL-CGT optimizes charge transfer efficiency up to $\sim 10^{13}$ cm$^{-2}$, contributing to unipolar transport and the enhanced FET ON-OFF ratio. {Second, the applied vertical electric field significantly amplifies the MAE of the CGT/CrOCl hybrid, establishing a high thermodynamic energy barrier for magnetization reversal based on the single domain nature of FL-CGT\cite{vervelaki2024visualizing}. } Consequently, the combined effects between an electrically enhanced MAE and interfacial hole doping manifest as the highly efficient, giant tunability of the coercive field.

\subsection*{Perspectives on substrate-boosted {spin transistor prototype}.} 

\begin{figure}[ht!]
\includegraphics[width=0.85\linewidth]{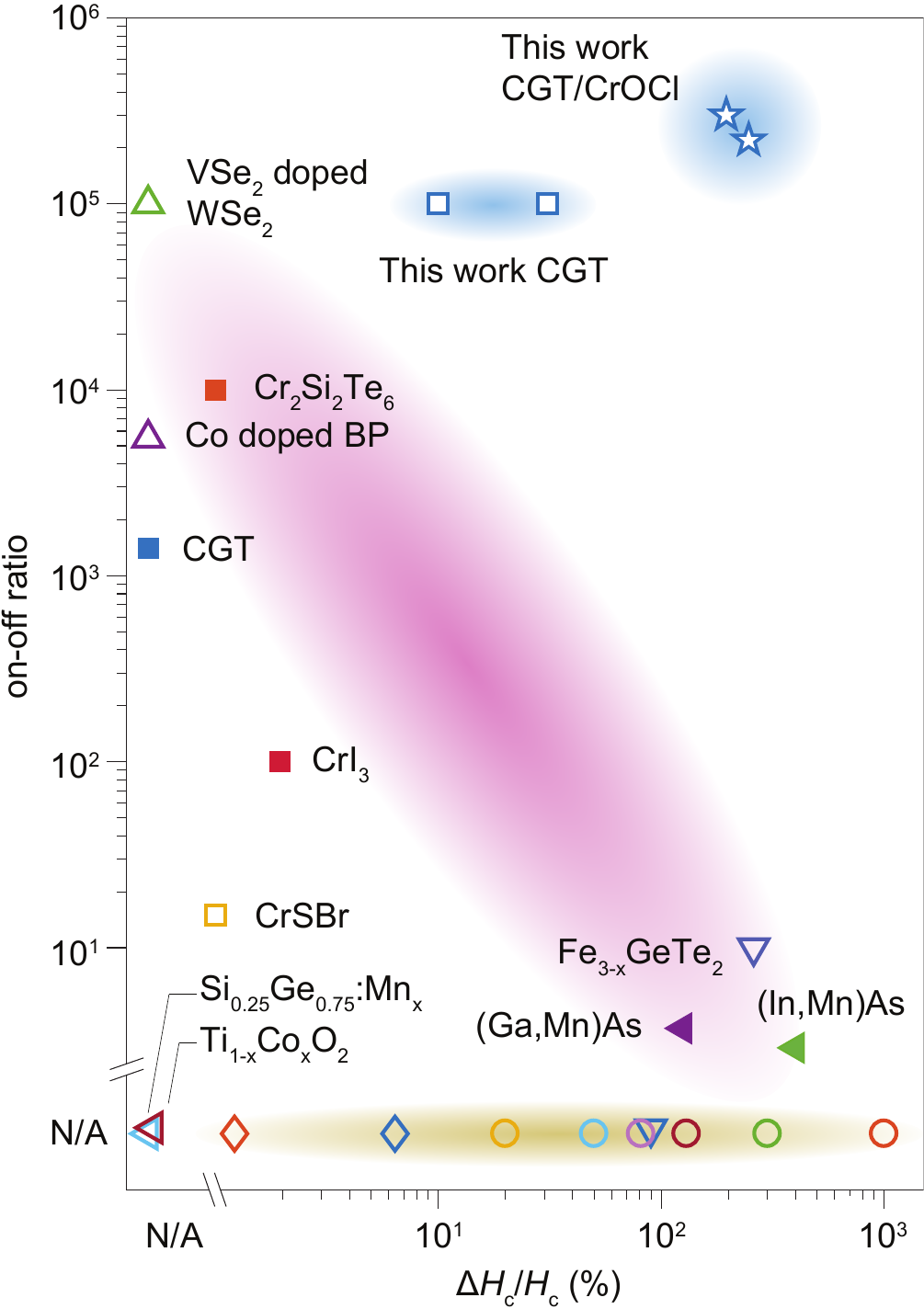}
\caption{\textbf{Perspectives on manipulation of charge and spin degrees of freedom in FMS-based FETs.} The diagram summarizes coercivity fields tunability and electrical FET ON-OFF ratios in several typical FMS-based FETs with using solid gates. Squares: 2D-FMS\cite{lin2016ultrathin,jiang2018controlling,Wang2018a,zhao2025doping}, left-pointing triangles: DMS\cite{ohno2000electric,chiba2003electrical,matsumoto2001room,wang2020high}, regular triangles: magnetic atom doped semiconductors\cite{yun2020ferromagnetic,fu2025electric}, inverted triangles: 2D ferromagnetic metals\cite{eom2023voltage,wang2025giant}, circles: ferromagnetic films\cite{wang2012electric,zhang2014electric,xue2019large,robbennolt2020magneto,xu2021voltage}, diamond: ferromagnetic insulators\cite{yu2019nonvolatile,al2025strain}. Data from devices in this work are included.
}
\label{fig:fig4}
\end{figure}

To evaluate the technological potential of the CGT/CrOCl heterostructure FETs, we benchmark our results against the broader landscape of state-of-the-art voltage-controlled FMS systems. Figure~\ref{fig:fig4} illustrates the relationship between the FET ON-OFF ratio and the relative coercive field modulation facilitated by solid-state gating across diverse magnetic classes, including DMSs, ferromagnetic films, magnet-doped semiconductors, and emerging 2D-FMSs. 

A comparative analysis of the compiled data reveals distinct performance trade-offs inherent to existing systems. Traditional DMSs (left-pointing triangles) exhibit pronounced $H_{\text{c}}$ tunability but are severely limited by low ON-OFF ratios. Ferromagnetic films, including both metallic (circles) and insulating (diamonds) variants, demonstrate negligible ON-OFF ratios due to screening effects or negligible intrinsic carrier densities. Conversely, magnet-doped semiconductors typically suffer from weak magnetic manipulation despite their semiconducting nature. Notably, our CGT/CrOCl devices occupy the top-right quadrant of the benchmark plot, simultaneously achieving a high ON-OFF ratio exceeding $10^5$ and a record-high $\Delta H_{\text{c}}/H_{\text{c}}$ surpassing 200\%.

This dual-parameter optimization represents a significant advancement over pristine FL-CGT, where the modulation efficiency is fundamentally constrained by limited carrier density control and weaker intrinsic anisotropy. Remarkably, our system outperforms both traditional DMS materials, such as $(\text{Ga,Mn})\text{As}$ and $(\text{In,Mn})\text{As}$, and contemporary 2D-FMS, including $\text{CrI}_3$, $\text{CrSBr}$, and $\text{Fe}_{3-x}\text{GeTe}_2$ within these parameters. Although the effective operating temperature currently remains low, the synergistic integration of CrOCl as both a dominant P-type dopant and an active pinning layer enables a degree of magnetic tunability previously elusive in single-component systems. This interface engineering approach establishes a versatile strategy for the simultaneous enhancement of electronic and magnetic performance in low-dimensional spintronic devices.

\vspace{5mm}

To conclude, we have demonstrated a transformative substrate boosting approach to overcoming the long-standing performance bottlenecks in 2D ferromagnetic semiconductors. By strategically implementing a FL-CGT/CrOCl vdW interface, we simultaneously engineer a high-performance P-type channel with superior ON-OFF ratio exceeding $10^5$ and achieve an unprecedented level of magnetic coercivity modulation over 200\%, effectively breaking the trade-off between electric gate control and magnetic tunability that plagues single-component systems. Another core significance of our work lies in the discovery of a gate-controlled crossover from critical power-law scaling to a dual power law. In the moderate gating regime, the proximity-induced exchange coupling follows a standard power-law behavior, inherited from the magnetic order of substrate. Heavy hole doping would drive the system into a regime of the second power scaling, {indicating the emergence of potential interface-stabilized magnetic states.} {While the current global $T_{\text{c}}$ is confined to low temperatures, the ability to independently tune charge transport and magnetic anisotropy via a multifunctional reservoir establishes a fundamental proof-of-concept. This substrate-boosting strategy provides a versatile model system for investigating fundamental interfacial phenomena and charge-spin correlation mechanisms across a broader class of low-dimensional magnetic and topological heterostructures.}

\bibliography{Substratebooster}
\bibliographystyle{naturemag}

\section*{Methods}
\noindent \textbf{Fabrication of 2D-FMS FETs.}
The device fabrication began with the preparation of Ti/Au$\sim$5/20 nm bottom gate electrodes on a Si/SiO$_2$ substrate using electron beam lithography (EBL) followed by electron beam evaporation. The vdW heterostructure was assembled using a hybrid dry transfer method. First, an h-BN flake was transferred via polypropylene carbonate (PPC) onto the bottom gate, followed by the lamination of a CrOCl flake exfoliated on polydimethylsiloxane (PDMS). Few-layered Cr$_{2}$Ge$_{2}$Te$_{6}$ flakes were exfoliated onto PDMS inside an argon-filled glovebox (O$_2$ and H$_2$O levels < 0.01 ppm) to prevent degradation and transferred onto the CrOCl. A top structure was formed by sequentially picking up an h-BN flake and two graphite strips using PPC, which was then transferred onto the Cr$_{2}$Ge$_{2}$Te$_{6}$/CrOCl/h-BN stack. Source and drain electrodes were defined by EBL and the top h-BN layer was etched by reactive ion etching (RIE) with SF$_6$ plasma (30 sccm, 50 W) to expose the graphite contact regions, followed by the deposition of Ti/Au$\sim$5/50 nm. Finally, the top gate electrode was defined by EBL, followed by the evaporation of Ti/Au$\sim$5/5 nm.

\vspace{3mm}
\noindent \textbf{Optical and Electrical measurements.} 
Temperature-dependent measurements were conducted in an optical cryostat (base temperature $\sim 6$\,K) equipped with vacuum electrical feedthroughs. Quasi-static direct-current characteristics, including transfer curves and I-V sweeps, were acquired using Keithley 2400 source meter units. The magnetic response was probed via polar MOKE microscopy using a linearly polarized semiconductor laser source (photon energy 1.94 eV). The Kerr rotation of the beam reflected at normal incidence was detected using a balanced photodiode bridge and standard lock-in techniques, as detailed in our previous reports\cite{Wang2018a}. {The coercive fields $H_c$, residue magnetization $M_r$, and saturation magnetization $M_s$ were extracted from the measured Kerr rotation loops by identifying the magnetic field that Kerr rotation crosses zero, the Kerr rotation at zero field, and the averaged Kerr rotation around $\pm50$ mT, respectively.} 

For room-temperature measurements, micro-Raman spectroscopy was performed using a Horiba LabRAM Odyssey system. Spectra were excited with a 532 nm single-frequency solid-state laser and collected through a $100\times$ objective lens to ensure high spatial resolution and signal-to-noise ratio. High-precision electrical characterizations were carried out using an Agilent B1500A Semiconductor device parameter analyzer integrated with a Cascade M150 probe station. 

\vspace{3mm}
\noindent \textbf{DFT and micromagnetic simulations.}
The electronic band structure and charge transfer calculations in this work were carried out by using first-principles density functional theory as implemented in the VASP code~\cite{vasp}. The projector augmented-wave (PAW) pseudopotential~\cite{paw} and the generalized gradient approximation (GGA) of Perdew-Burke-Ernzerhof (PBE) functional were used for electron-ion and electron-electron exchange-correlation interaction, respectively. GGA+U was used~\cite{GGAU} for the transition metal Cr with the Coulomb repulsion U of 0.8 eV~\cite{Wang2018a}. The Brillouin zone of the super cell of CGT/CrOCl hybrid was sampled by 8$\times$4$\times$1 k mesh. The electronic kinetic energy cutoff of 450 eV was adopted for the plane-wave basis and 1 $\times$ 10$^{-6}$ eV for the electronic self-consistency criterion. Hybrid structure was optimized using the conjugate gradient method, until none of the residual Hellmann-Feynman forces exceeded 10$^{-2}$ eV/\AA. Spin-orbit coupling and magnetocrystalline anisotropy energy (MAE) were calculated by relativistic pseudopotentials derived from an atomic Dirac-like equation. Micromagnetics simulations were performed by using an atomistic spin model based on the Heisenberg Hamiltonian as implemented in the object oriented micromagnetic framework (OOMMF) software package~\cite{oommf}.

\section*{Data Availability}
The data that support the findings of this study will be available on Zenodo when published.

\section*{Code Availability}
The code that support the findings of this study are available upon reasonable request to the corresponding authors.

\section*{Acknowledgement}
The authors thank Zheng Han for fruitful discussions. This work was funded by the National Key R\&D Program of China (Nos. 2025YFA1411100, 2022YFA1203900) and {the National Natural Science Foundation of China (NSFC) (Grant Nos. 12674227, 62375160, 62274180, 62504144, 92264201, U24A6002, 52471203, 52031014)}. T.Z. acknowledges the Research Project Supported by Shanxi Scholarship Council of China. J.W and J.C. acknowledge the support from National Natural Science Foundation of Shandong Province (ZR2025ZD41, TSQN202306059). R.C. acknowledge the support from Interdisciplinary Research Program of Natural Science of Hebei University (No.DXK202212) and Central Government Guided Local Science and Technology Development Fund Project (226Z1703G). Hanwen W. acknowledges support from National Natural Science Foundation of Liaoning Province, China (Grant No. 2024JH3/50100022). K.W. and T.T. acknowledge support from the CREST (JPMJCR24A5), JST and World Premier International Research Center Initiative (WPI), MEXT, Japan.

\section*{Author Contributions}
T.Z. conceived the experiment and supervised together with X.X., R.C., J.C., and J.W. for the overall project. H.W., R.C. carried out device fabrications, with participation from W.P., G.Z., Xiaoyan L., Xiaoxi L. and Hanwen W.. H.W., W.P. and T.Z. conducted DC electrical measurements. W.P. and T.Z. performed the MOKE measurements at cryogenic temperatures. T.Y., Y.T., X.K.,and W.G. performed first-principles calculations. K.W. and T.T. provided high quality h-BN bulk crystals. T.Z., J.W., J.C., T.Y., H.W. and X.X. analyzed the experimental data. The manuscript was written by T.Z. with discussions and inputs from all authors.

\section*{Competing Interests}
The authors declare no competing interests.

\end{document}